\documentclass[twocolumn,showpacs,preprintnumbers,nofootinbib,amsmath,amssymb]{revtex4}

\usepackage{amsfonts} % para incluir fonts y
\usepackage{amssymb}  % simbolos de la AMS
\usepackage{graphicx} % standard LaTeX graphics tool
\usepackage{amsmath}  % for including eps-figure files
\usepackage{amsmath}  % paquete matemÃ¡tico
\usepackage[latin1]{inputenc}
\usepackage{color}
\usepackage{natbib}

\begin{document}
\title{Anisotropic equation of state of charged and neutral vector boson gases in a constant magnetic field. Astrophysical implications}
\author{G. Quintero Angulo\footnote{gquintero@fisica.uh.cu}}
\affiliation{Facultad de F{\'i}sica, Universidad de la Habana,\\ San L{\'a}zaro y L, Vedado, La Habana 10400, Cuba}
\author{A. P\'erez Mart\'{\i}nez\footnote{aurora@icimaf.cu}}
\affiliation{Instituto de Cibern\'{e}tica, Matem\'{a}tica y F\'{\i}sica (ICIMAF), \\
 Calle E esq a 15 Vedado 10400 La Habana Cuba\\}
\author{H. P\'erez Rojas\footnote{hugo@icimaf.cu}}
\affiliation {Instituto de Cibern\'{e}tica, Matem\'{a}tica y F\'{\i}sica (ICIMAF), \\
 Calle E esq a 15 Vedado 10400 La Habana Cuba\\}

%\presentaddress{Present address. Present address. Present address}

\begin{abstract}%
We obtain the pressures and equations of state (EoS) of charged and neutral vector boson gases in a constant magnetic field. The axial symmetry imposed to the system by the field splits the pressures in the parallel and perpendicular directions along the magnetic axis, and this leads to anisotropic equations of state. The values of pressures and energy densities are in the order of those of Fermi gases in compact objects. This opens the possibility to the existence of magnetized boson stars. Under certain conditions, the perpendicular pressure might be negative imposing a bound to the stability of the star. Other implications of negative pressures are also discussed.
\end{abstract}


\maketitle

\section{Introduction}\label{intro}

Compact objects have been extensively studied considering gravity counterbalanced by the degenerate Fermi pressure. Since the decade of 1960s has been contemplated the possibility of cold Bosons Stars, which compensate gravity with the Heinsenberg pressure \cite{PhysRev.187.1767, PhysRevD.38.2376}. These works were considered merely as an academic issue because a noninteracting Bose gas at zero temperature leads to obtain denser objects with masses and radius smaller than those typical of fermions stars \cite{PhysRevD.38.2376}.
Nevertheless, the discovery of Bose-Einstein condensation in lab \cite{Anderson198} has triggered a great interest in these self-gravitating objects. On the other hand, models at finite temperature and/or with the inclusion of interaction gives maximum masses and radii comparable with neutron stars \cite{Takasugi1984,latifah2014bosons}. Besides, the existence of mixed fermion-bosons stars is an open possibility. In this frame, vector bosons could be useful in explaining the strong magnetic fields shown by these objects, since they are known to sustain its own magnetic field \cite{Yamada,ASNA:ASNA201512243,us}.

We study the anisotropic pressures and the equations of state (EoS) for the charged and the neutral vector boson gas (CVBG and NVBG respectively) in a constant magnetic field starting from the spectra given by Proca theory. The EoS are the first step to study the structure equations to obtain observables: maximum masses and radii of magnetized bosons stars.
%Since we are focussed on possible astrophysical applications we deal with systems densities in the order of $10^{34}cm^{-3}$.

In Section~\ref{TD}, we present the spectrum and the thermodynamical potential of charged and neutral vector boson in a constant magnetic field. In Section~\ref{Pressures} the anisotropic pressures are discussed while Section~\ref{EoS} is devoted to the EoS. Concluding remarks are given in Section~\ref{concl}.

\section{Thermodynamical properties}\label{TD}
The energy spectra for the CVBG \cite{PhysRevD.60.033003} and the NVBG \cite{us} in a constant magnetic field $\textbf{B} = (0,0,B)$ are:
\begin{eqnarray}
\varepsilon^{ch}(p_3,n)&=&\sqrt{m^2+p_3^2+ (2n + 1 -2S)q B}, \\[10pt]
\varepsilon^{n}(p_3,p_{\perp})&=&\sqrt{m^2+p_3^2+p_{\perp}^2-2\kappa s B\sqrt{p_{\perp}^2+m^2}},
\end{eqnarray}
where $s=0, \pm 1$ are the spin eigenvalues, $n = 0, 1, 2...$ labels the Landau levels, $p_3$ is the momentum component along the magnetic field and $p_{\perp}$ is the momentum component perpendicular to the magnetic field. In $\varepsilon^{ch}(p_3,n)$, $p_{\perp}$ has been replaced by its quantized values in terms of $n$.

The ground states of the vector boson gases are obtained from the spectra by setting $p_3 = 0$, $p_{\perp} = 0$, $n=0$ and $s = 1$. In both cases
\mbox{$\varepsilon^{ch,n}=\sqrt{m^2 - q B} = m \sqrt{1-b}$}, with $b=B/B^{ch,n}_c$ being $B^{ch,n}_c$ the values of the magnetic field for which $\varepsilon^{ch,n}=0$. The critical fields are $B^{n}_c = m / 2 \kappa$ for the NVBG and $B^{ch}_c = m^2 / q$ for the CVBG. For numerical calculations we suppose the charged bosons to be two paired electrons and have a mass $m = 2 m_e$ ($m_e$ is the electron mass), and a charge $q = 2 e$ ($e$ is the electron charge). For the neutral bosons we use a positronium gas parameters, whit mass $m = 2 m_e$ and magnetic moment $\kappa = 2\mu_B $ ($\mu_B$ is the Bohr magneton).

The spectra allows us to obtain the thermodynamical potentials. In general we can write them in the form $\Omega^{ch,n} = \Omega^{ch,n}_{st} + \Omega^{ch,n}_{vac}$, were $\Omega^{ch,n}_{vac}$ stand for the vacuum contributions and is only $b$ dependent, while $\Omega^{ch,n}_{st}$ stand for the particles and depend on $T$ and $b$.

For the CVBG in the low temperature limit ($T \ll m$), $\Omega^{ch}_{st}$ \cite{PhysRevD.60.033003} is
\begin{equation}\label{Potencialstch}
\Omega^{ch}_{st} =
\left \{  \begin{array}{ll}
- \frac{3 (\varepsilon^{+})^{3/2}}{(2 \pi)^{3/2} \beta^{5/2}} Li_{5/2}(e^{\beta(\mu - \varepsilon^{+})}), & {\bf WF}\\
- \frac{m^2 b (\varepsilon)^{1/2}}{3^{3/2} \pi^{5/2} \beta^{3/2}} Li_{3/2}(e^{\beta(\mu - \varepsilon)}), & {\bf SF}
\end{array} \right.
\end{equation}

In Eq.~(\ref{Potencialstch}) $\mu$ is the chemical potential, $\beta$ is the inverse temperature, $\varepsilon^{+} = m \sqrt{1+b}$ and $Li_{k}(x)$ is the polylogarithmic function of order $k$. The condition $T = m b$ separates the weak field ({\bf WF}) $T > m b$  from the strong field ({\bf SF}) $T < m b$ region.

Vacuum contribution to the thermodynamical potential of the CVBG gas after regularization can be written as
\begin{widetext}
\begin{equation}\label{Potencialvacch}
\Omega^{ch}_{vac}  =
\left \{
\begin{array}{ll}
-\frac{3 m^4}{64 \pi^2} \left\{ 12 b^2-(1-b)^2 \log(1-b) + 2(1+2b+5b^2) \log(1+b)-(1-3b)^2\log(1+3b)\right\}, & \, {\bf WF} \\
-\frac{3 m^4 b}{16 \pi^2}\left\{ -\frac{2b^2}{1+b} +\frac{1}{2} \left[(1-b) \log(1-b) -2(1+b)\log(1+b) +(1+3b) \log(1+3b)\right] \right\}, & \, {\bf SF}.
\end{array}
\right.
\end{equation}
%\end{spaneqn}
The statistical and vacuum contributions to the NVBG thermodynamical potential in the low temperature limit read \cite{us}
%\begin{spaneqn}
\begin{eqnarray}\label{Potencialn}\nonumber
\Omega^{n}_{st}&=&-\frac{(\varepsilon^{n})^{3/2}}{2^{1/2} \pi^{5/2} \beta^{5/2} (2-b)} Li_{5/2}(e^{\beta \mu^{\prime}}),\\ [10pt]
\Omega^{n}_{vac}&=&-\frac{m^4}{288 \pi} \left \{ b^2(66-5 b^2)-3(6-2b-b^2) (1-b)^2 \log(1-b)-3(6+2b-b^2)(1+b)^2 \log(1+b) \right \}.
\end{eqnarray}
\end{widetext}

Eqs.~(\ref{Potencialn}) are valid for any field value because in the neutral case it is not necessary to consider separately the weak/strong field regimes (\cite{us}).

The magnetization for the charged ($M^{ch}$ \cite{PhysRevD.60.033003,ROJAS1996148}) and the neutral ($M^{n}$ \cite{us}) gas read
\begin{eqnarray}\label{mag}\nonumber
M^{ch} &=&
\left \{
\begin{array}{ll}
\frac{7 q b m^{3/2}}{4 (2 \pi)^{3/2} \beta^{1/2}} e^{\beta(\mu - m)}, &  \, {\bf WF}\\
\frac{q}{2 \varepsilon(b)} N = \frac {q}{2 m \sqrt{1-b}} N, & \, {\bf SF},
\end{array}
\right.\\[10pt]
M^{n} &=& \frac{\kappa m}{\varepsilon^n} N = \frac{\kappa}{\sqrt{1-b}} N.
\end{eqnarray}
with $N$ the density of particles. It is interesting to note that the magnetization of the CVBG in the strong field regime coincides with the one of the NVBG (for the charged gas we can define a magnetic moment equal to $q/2m$). Thus, their magnetic properties at strong field values are expected to be the same.

\section{Anisotropic pressures}\label{Pressures}
The pressures for the magnetized CVBG and NVBG are \cite{chaichian2000quantum}
\begin{eqnarray}\label{P}\nonumber
P^{ch,n}_3 &=&-\Omega^{ch,n} = -\Omega^{ch,n}_{st} -\Omega^{ch,n}_{vac},\\[10pt]
P^{ch,n}_{\perp} &=&-\Omega^{ch,n}-B M^{ch,n} = P^{ch,n}_3-B M^{ch,n}.
\end{eqnarray}
where $P^{ch,n}_3$ ($P^{ch,n}_{\perp}$) is the pressure in the parallel (perpendicular) direction with respect to the magnetic axis. The splitting of the pressure is a consequence of the axial symmetry imposed by the magnetic field.

\begin{figure}[h!]
\includegraphics[width=0.8\linewidth]{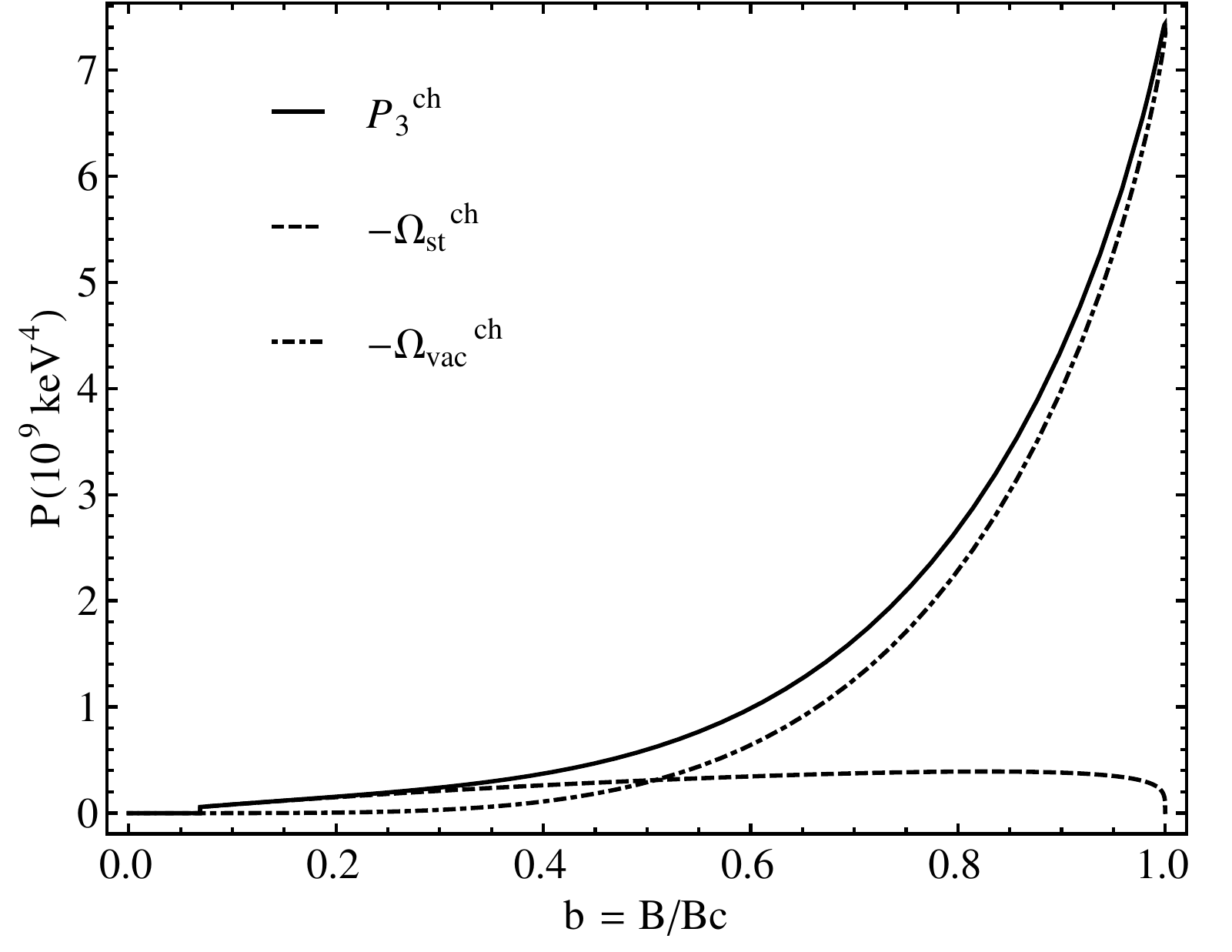}
\includegraphics[width=0.8\linewidth]{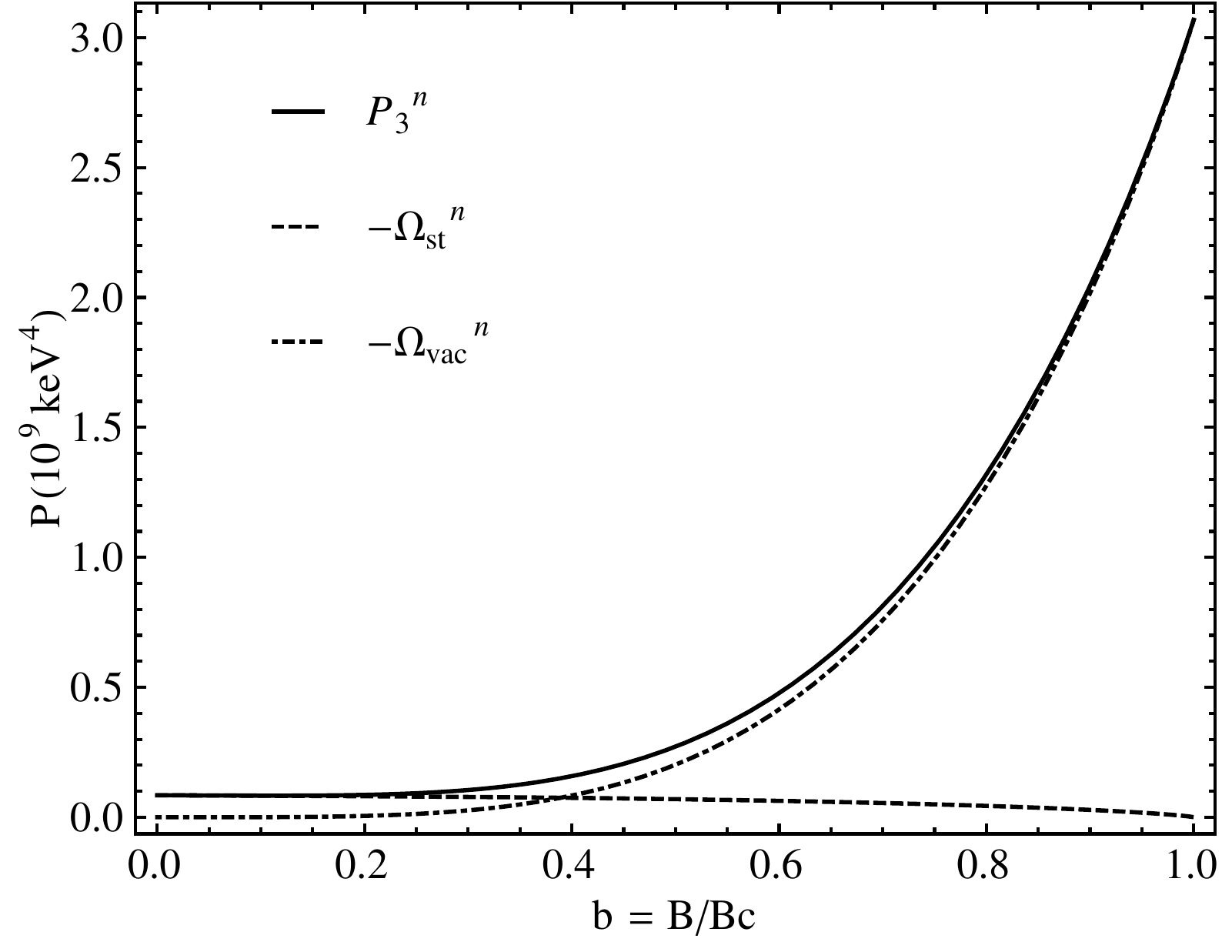}
{\caption{The parallel pressure $P_3$ as a function of the magnetic field for the CVBG (upper panel) and the NVBG (lower panel); the statistical and the vacuum contributions to the pressure are also plotted in dashed and dot-dashed lines. We used $T = 8 \times 10^8$ K and $N = 10^{34}$ cm$^{-3}$.\label{fig1}}}
\end{figure}

Fig.~1 shows $P^{ch}_3$ (upper panel) and $P^{n}_3$ (lower panel) for $T = 8 \times 10^8$ K and $N = 10^{34}$ cm$^{-3}$. We also plotted the statistical ($-\Omega^{ch,n}_{st}$) and the vacuum ($-\Omega^{ch,n}_{vac}$) pressures in dashed and dot-dashed lines. In general, the behaviour of $P^{ch}_3$ and $P^{n}_3$ is the same, except for the jump observed in $P^{ch}_{3}$ (Fig.~\ref{fig1}, upper panel) around $b=0.05$ that corresponds to the limit between the weak/strong field approximations $T = m b$. A similar jump can be also found in the perpendicular pressure of the CVBG (Fig.~2, upper panel).

When $b=0$, the values of $P^{ch,n}_3$ coincide with those of theirs statistical parts ($-\Omega^{ch,n}_{st}$), but as the field grows the parallel pressures increases and approaches theirs vacuum contributions ($-\Omega^{ch,n}_{vac}$), while the statistical parts decreases and goes to zero for $b=1$ (increasing the field drives the system to BEC \cite{us}). Since $-\Omega^{ch,n}_{vac}$ are temperature independent, a change in temperature within the $T \ll m$ limit will not affect substantially the parallel pressures.

Fig.~2 shows the perpendicular pressure $P^{ch,n}_{\perp}$ for the CVBG (upper panel) and the NVBG (lower panel) for $T = 8 \times 10^8$ K and several values of the particle density $N$. For both gases the perpendicular pressures starts from a positive value in $b=0$,  decreases with $b$ and eventually becomes negative. This is because the main contribution to $P^{ch,n}_{\perp}$ comes from the magnetic pressure terms $-M^{ch,n} B$ which are always negative and diverge in the critical field. Whether the perpendicular pressure is positive or not depend on the field, the temperature and the particle density \cite{us}. A negative perpendicular pressure pushes the particles inward to the magnetic field axis, while they are pushed outward in the direction of the field by the parallel pressure that is always positive. This kind of instability is known as transversal magnetic collapse \cite{chaichian2000quantum} and might be relevant in the description of ejection of mass and radiation out of astronomical objects \cite{ASNA:ASNA201512243}.

\begin{figure}[h!]
\includegraphics[width=0.8\linewidth]{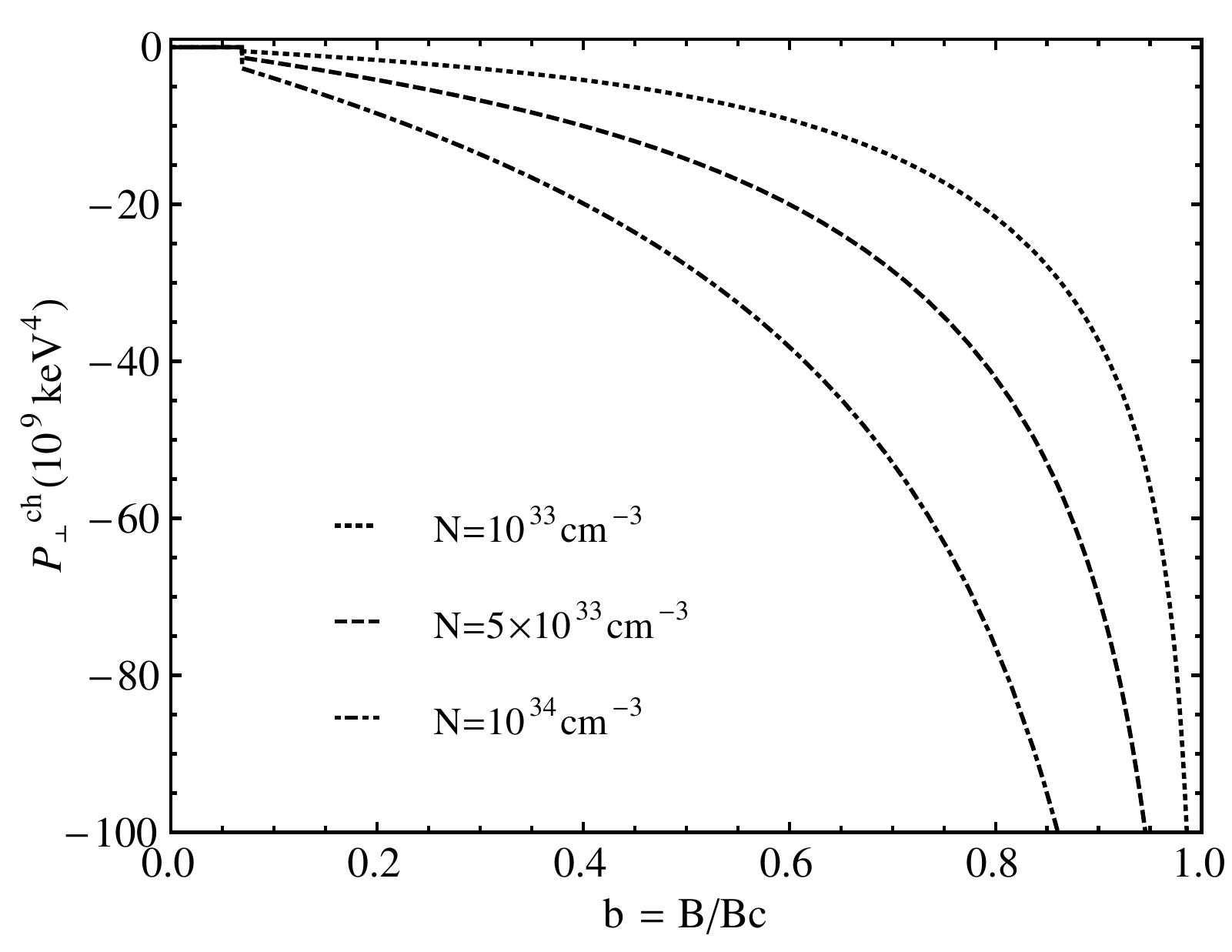}
\includegraphics[width=0.8\linewidth]{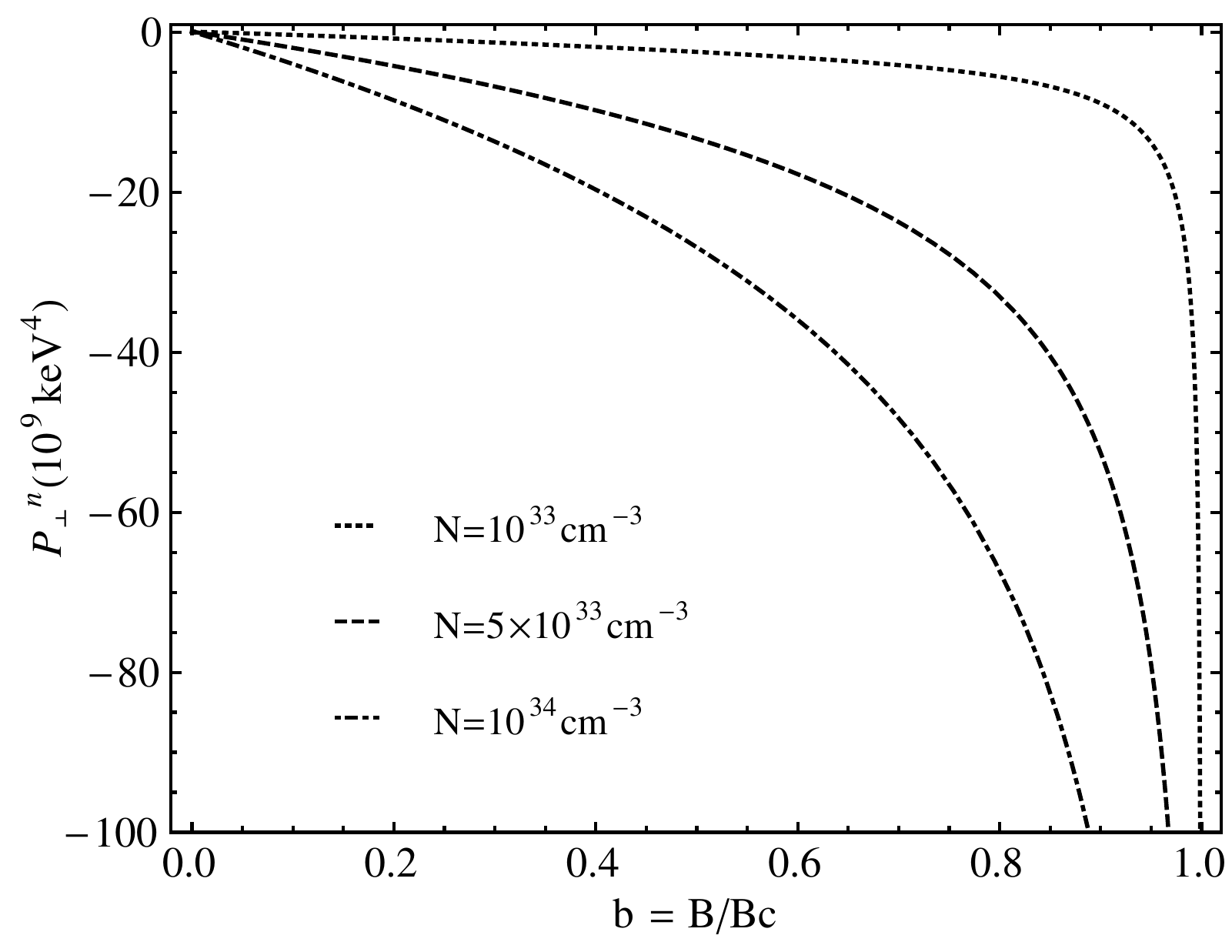}
{\caption{The perpendicular pressure $P_{\perp}$ as a function of the magnetic field for the charged (upper panel) and the neutral (lower panel) vector boson gas for $T = 8 \times 10^8$ K and several values of the particle density $N$.\label{fig3}}}
\end{figure}

The positives values of the pressures are in the order of those of an electron gas in a white dwarf ($10^{-3}$ MeV$^4$). Moreover, if we consider in our calculations $\rho$ mesons or bosons resulting from the pairing of two neutrons, we get that their positive pressures are around $10^{8}$ MeV$^4$  and $10^{9}$ MeV$^4$ respectively. This values are in the same order of magnitude to those exerted by a gas of neutrons ($10^{9}$ MeV$^4$) or quarks ($10^{8}$ MeV$^4$) in a compact star. In consequence, a gas of magnetized neutral or charged vector bosons might oppose gravity and maintain stability in a pure -or mixed- boson star.

\section{Equations of State}\label{EoS}
Starting from the thermodynamic potentials we can compute the energy of the vector boson gases through it definition \mbox{$E = -T \partial \Omega/\partial T +\Omega+\mu N$}. We get
\begin{eqnarray}\nonumber
E^{ch} \hspace{-2pt} &=& \hspace{-2pt} \left \{ \hspace{-2pt}
\begin{array}{ll}
\varepsilon^+ N +\frac{5 }{2} \Omega^{ch}_{vac} \hspace{-2pt} + \hspace{-2pt} T N \frac{\partial \mu^{ch}}{\partial T} \hspace{-0.7pt} - \hspace{-0.7pt}\frac{3}{2} \Omega^{ch}, & \, {\bf WF}\\
\varepsilon N + \frac{3}{2} \Omega^{ch}_{vac} \hspace{-2pt} + \hspace{-2pt} \frac{(2 \pi-1)\mu^{\prime ch}}{2 \pi} N \hspace{-0.7pt} - \hspace{-0.7pt} \frac{1}{2} \Omega^{ch} , & \, {\bf SF},
\end{array}
\right. \\[10pt] \label{energy}
E^n  &=& \varepsilon^n N + \Omega^n_{vac}- \frac{3}{2} \Omega^n_{st} - \frac{\partial \mu^{\prime n}}{\partial \beta} N,
\end{eqnarray}

with
\begin{eqnarray}\nonumber
\mu^{\prime ch} &=& -\frac{q m b T^2 \varepsilon(b)}{8 \pi^2 N^2},\\ [10pt] \nonumber
\mu^{ch} \hspace{-0.7pt}&=&\hspace{-0.7pt} \left( \frac{N}{3}\hspace{-1pt} \left( \frac{2 \pi}{\varepsilon^+}\right)^{3/2}\hspace{-0.7pt}- \hspace{-0.7pt}Li_{3/2}(e^{- \beta \varepsilon^+})\right) \frac{T}{Li_{1/2}(e^{- \beta \varepsilon^+})},\\[10pt] \nonumber
\mu^{\prime n} &=& -\frac{\zeta(3/2)T}{4 \pi} \left ( 1- \left(\frac{T_{cond}}{T} \right)^{3/2} \right )\Theta(T-T_{cond}),\\[10pt] \nonumber
T_{cond} &=& \frac{1}{\varepsilon^n} \left ( \frac{2^{1/2} \pi^{5/2} (2-b) N}{\zeta(3/2)}\right)^{2/3}.\nonumber
\end{eqnarray}

Combining Eqs.~(\ref{energy}) with Eqs.~(\ref{P}) we obtain the equations of state for the charged
\begin{eqnarray}\label{Pch}
\hspace{-6pt} P^{ch}_3 \hspace{-2pt} &=& \hspace{-2pt} \left \{ \hspace{-2pt} \begin{array}{ll}
\frac{2}{3}\left( E^{ch} \hspace{-2pt} - \hspace{-2pt} \varepsilon^{+} N \hspace{-2pt} + \hspace{-2pt} \frac{5}{2} \Omega^{ch}_{vac}\hspace{-2pt} + \hspace{-2pt} T N \frac{\partial \mu^{ch}}{\partial{T}}\right)\!, &  \, {\bf WF}\\
2\left( E^{ch} \hspace{-2pt} - \hspace{-2pt} \varepsilon^{ch} N \hspace{-2pt} + \hspace{-2pt} \frac{3}{2} \Omega^{ch}_{vac} \hspace{-2pt} + \hspace{-2pt}\frac{(2 \pi-1)\mu^{\prime ch}}{2 \pi} \hspace{-2pt} N\right)\!, & \, {\bf SF},
\end{array} \right.
\\[10pt] \nonumber \label{Ech2}
\hspace{-6pt} P^{ch}_{\perp} \hspace{-2pt}  &=& \hspace{-3pt}  \left \{ \hspace{-3pt}  \begin{array}{ll}
 \frac{2}{3}\left( E^{ch} \hspace{-3pt} - \hspace{-2pt} \varepsilon^{+} N \hspace{-2pt}  + \hspace{-2pt} \frac{5}{2} \Omega^{ch}_{vac}\hspace{-2pt} + \hspace{-2pt}  T \hspace{-2pt} N \hspace{-1pt} \frac{\partial \mu^{ch}}{\partial{T}}\right)\hspace{-3pt}- \hspace{-3pt} M^{\hspace{-1pt} ch} \hspace{-2pt} B, & \, {\bf WF} \\ \nonumber
 2 \left(E^{ch} \hspace{-3pt}  - \hspace{-2pt} \varepsilon N \hspace{-2pt}  + \hspace{-2pt}  \frac{3}{2} \Omega^{ch}_{vac} \hspace{-2pt}  + \hspace{-2pt} \frac{(2 \pi -1)\mu^{\prime ch}}{2 \pi} \hspace{-2pt}  N \hspace{-2.5pt}\right) \hspace{-3pt}- \hspace{-3pt} M^{\hspace{-1pt} ch} \hspace{-2pt} B, & \, {\bf SF},
\end{array} \right.
\end{eqnarray}
and the neutral gas
\begin{eqnarray}\label{Pn}
P^{n}_3 &=& E^n-(\varepsilon^n(b) N - \frac{5}{2} \Omega^n_{st} - \frac{\partial \mu^{\prime n}}{\partial \beta} N),\\[10pt] \nonumber
P^{n}_{\perp} &=& \hspace{-2pt} E^n-(\varepsilon^n(b) N -\hspace{-2pt} \frac{5}{2} \Omega^n_{st} -\hspace{-2pt} \frac{\partial \mu^{\prime n}}{\partial \beta} N + \hspace{-2pt} M^n B).
\end{eqnarray}

\begin{figure}[h!]
\includegraphics[width=0.8\linewidth]{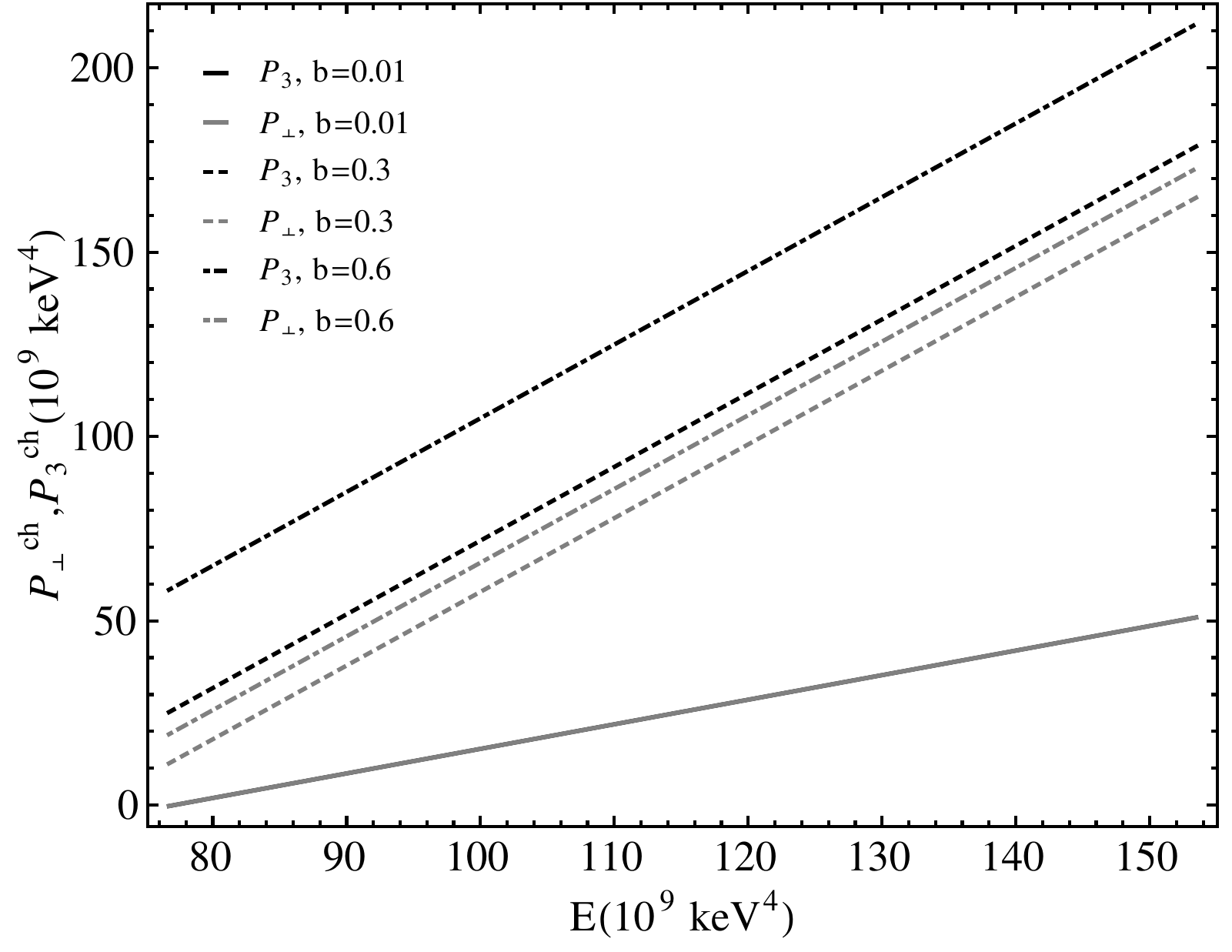}
\includegraphics[width=0.8\linewidth]{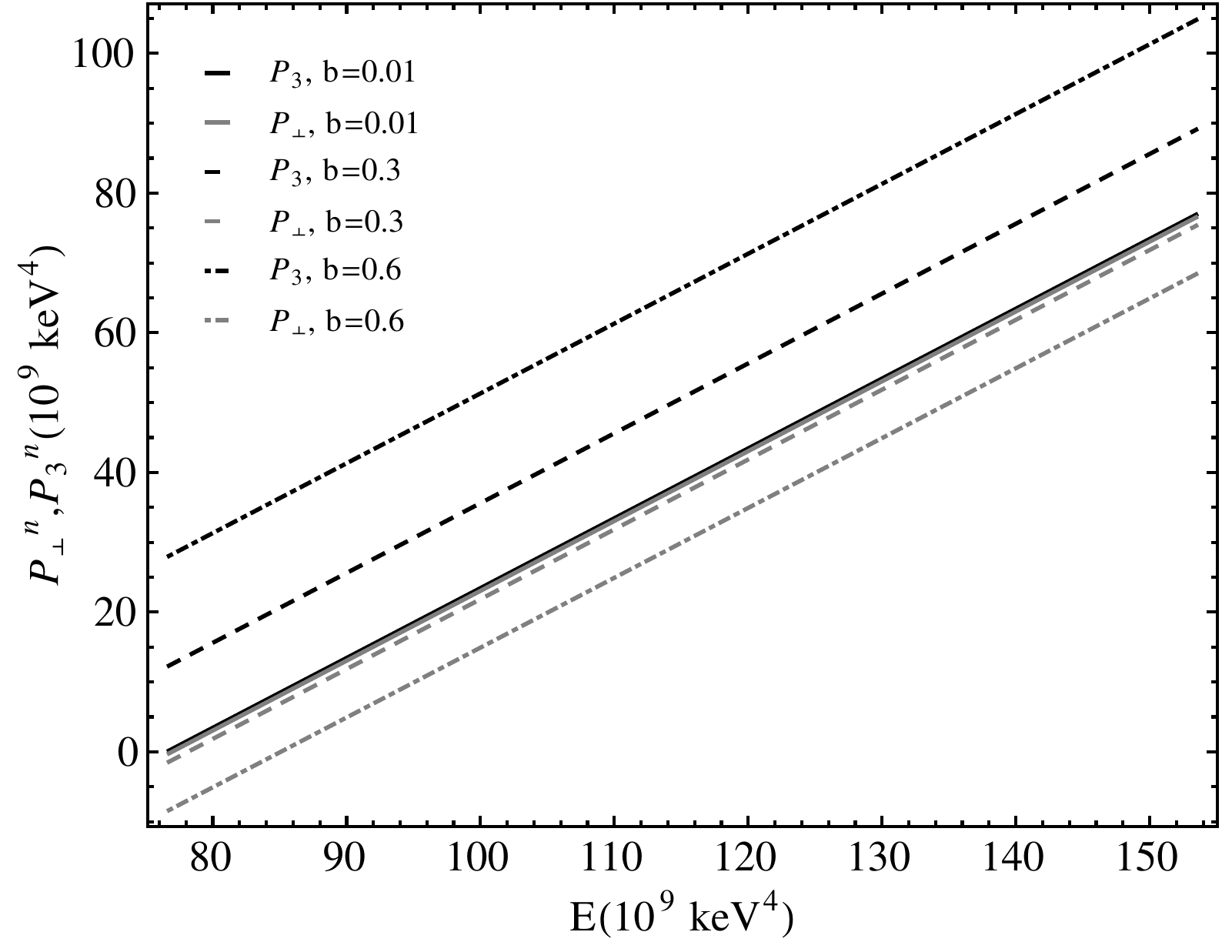}
{\caption{The equations of state for the charged (upper panel) and the neutral vector boson gas (low panel). $T = 8 \times 10^8$ K and \mbox{$N=10^{34}$ cm$^{-3}$}.\label{fig3}}}
\end{figure}
In Fig.~3 the EoS ($P_3, P_{\perp}$ vs $E$) are depicted for \mbox{$T = 8 \times 10^8$ K and $N=10^{34}$ cm$^{-3}$}. Upper panel corresponds to the charged gas while lower panel to the neutral. Again both gases behaves similarly, the pressures being linear functions of the energy. When the field is weak (solid lines) the difference between the parallel (in black) and the perpendicular  (in gray) pressure is negligible an the systems can be described by an unique equation of state like in the zero field case. As the field grows the pressures split out and the anisotropy starts to be significant for the EoS. For a fixed value of the field, the EoS for the perpendicular pressures are softer that the corresponding parallel one. The change of slope from the $b=0.01$ curve to the $b=0.3$ and $b=0.6$ that occurs for the CVBG (upper panel of Fig.~3) is due to the change from the weak to the strong field approximation.

\section{Conclusions}\label{concl}
We have studied the anisotropic pressures and EoS of the NVBG and the CVBG in the presence of a constant magnetic field and found that both gases behave similarly. The parallel pressure is always positive and increases with the magnetic field, while the perpendicular decreases and reaches negative values eventually, as happens for neutral and charged fermions systems in presence of magnetic field \cite{chaichian2000quantum}.

The positive values of the pressures are high enough to oppose gravity and maintain the stability in a boson star. Following this direction, our next step will be to study the solution of the structure equations with our EoS in order to obtain maximum values of masses and radii of magnetized Boson stars. The negative parallel pressures imposes a bound for the magnetic field of a stable compact star that depends on the mass, the temperature and the density of the particles considered, but the possibility of having a magnetic collapse might play a major role in understanding the ejection of mass and radiation out of astronomical objects.

\section{Acknowledgements}
The work of G.Q.A, A.P.M. and H.P.R. have been supported under the grant CB0407 and the ICTP Office of External Activities through NET-35.

\end{document}